\pdfoutput=1

\documentclass[prl,superscriptaddress,amsmath,amssymb,twocolumn]{revtex4-1}

\usepackage{graphicx,bm,amsmath}
\usepackage[usenames,dvipsnames]{xcolor}
\usepackage[colorlinks,bookmarks=false,citecolor=NavyBlue,linkcolor=Red,urlcolor=blue]{hyperref}
\usepackage[bbgreekl]{mathbbol}
\usepackage{xcolor}
\usepackage[normalem]{ulem}
\usepackage{braket}
\usepackage{bm}
\usepackage{esint}

\def\doi{http://dx.doi.org/}

\newcommand{\be}{\begin{equation}}
\newcommand{\ee}{\end{equation}}
\newcommand{\bea}{\begin{eqnarray}}
\newcommand{\eea}{\end{eqnarray}}

\newcommand{\titleinfo}{Thermalisation of a trapped one-dimensional Bose gas via diffusion}

\begin{document}

%%%%%%%%%%%%%%%%%%%%%%%%%%%%%%%%%%%%%%%%%%%%%%%%%%%%%%%%
\title{\titleinfo}
%%%%%%%%%%%%%%%%%%%%%%%%%%%%%%%%%%%%%%%%%%%%%%%%%%%%%%%%

\author{Alvise Bastianello}
\affiliation{Institute for Theoretical Physics, University of Amsterdam, Science Park 904, 1098 XH Amsterdam, The Netherlands}

\author{Andrea De Luca}
\affiliation{Laboratoire de Physique Th\'eorique et Mod\'elisation (CNRS UMR 8089), Universit\'e de Cergy-Pontoise, F-95302 Cergy-Pontoise, France}    

\author{Benjamin Doyon}
\affiliation
{Department of Mathematics, King's College London, Strand WC2R 2LS, London, U.K.}

\author{Jacopo De Nardis}
\affiliation{Department of Physics and Astronomy, University of Ghent, Krijgslaan 281, 9000 Gent, Belgium}

\begin{abstract}
For a decade the fate of a one-dimensional gas of interacting bosons in an external trapping potential remained mysterious. We here show that whenever the underlying integrability of the gas is broken by the presence of the external potential, the inevitable diffusive rearrangements between the quasiparticles, quantified by the diffusion constants of the gas, eventually lead the system to thermalise at late times.  We show that the full thermalising dynamics can be described by the generalised hydrodynamics with diffusion and force terms, and we compare these predictions with numerical simulations. Finally, we provide an explanation for the slow thermalisation rates observed in numerical and experimental settings: the hydrodynamics of integrable models is characterised by a continuity of modes, which can have arbitrarily small diffusion coefficients. As a consequence, {the approach to thermalisation can display pre-thermal plateau and relaxation dynamics with long polynomial finite-time corrections.  }
\end{abstract}

\maketitle

In the past two decades, low-temperature gases of bosonic or fermionic atoms emerged as the best experimental platform where to study many-body phenomena and where to probe their non-equilibrium dynamics \cite{Langen2015,Nichols2018,2005.09549}. A great deal of attention has been dedicated to the problem of thermalisation in isolated systems, such as a single gas of bosonic atoms interacting via contact repulsion. The experiment of 2007 dubbed ``quantum Newton cradle'' \cite{Kinoshita2006} represented a turning point. The apparent lack of thermalisation at late times for a non-equilibrium one-dimensional cold atomic gas in an external trap started numerous research directions on the role of the underlying integrability in the isolated dynamics. It was then later clarified that integrable models do not thermalise to standard Gibbs ensembles, but to Generalises Gibbs ones \cite{rigol2007relaxation,Calabrese_2016,Langen2015}, where entropy is maximised given the constraints of all local and quasi-local conserved quantities. As the integrability is weakly broken by the external trap, one expects therefore to crossover from a fast relaxation to a generalised Gibbs state, to a slow pre-thermal decay to the final thermal ensemble. In the past years, the study of pre-thermalisation dynamics has received many contributions, and for homogeneous systems, a quite comprehensive understanding has now been achieved~\cite{Gring2012,Bertini2015,PhysRevB.94.245117,PhysRevLett.119.010601,Langen2013,Langen2016,Mallayya2019,PhysRevB.101.180302,Durnin2020,PhysRevResearch.2.022034,2006.13891}. However, a full description of the (pre) thermalisation mechanism in inhomogenous systems is still lacking. While a recent experimental work showed that multiple one-dimensional gases coupled to each other or where transitions to higher dimension are allowed, do indeed thermalise at late times  \cite{PhysRevX.8.021030,Moller2020}, it remains to be understood what is the fate of a single isolated one-dimensional gas.

In this letter, we show that in the presence of an external force and of diffusive spreading, a single gas described by an integrable Hamiltonian and by an external trapping potential does thermalise, due to the interplay of the external force that breaks integrability and the diffusion spreading that redistributes the quasiparticle momenta and increases the local thermodynamic entropy. Previous works have studied the hydrodynamics of an integrable gas in the presence of the external force but without accounting for diffusive terms~\cite{doyon2017note, CauxCradle,CaoCradle,PhysRevLett.123.130602,PhysRevLett.122.240606}. Here, we show that diffusive corrections play an essential role in the late-time dynamics. %Analogously to the hydrodynamics of a normal fluid, the gas reacts to the external gradients by means of its convective currents,  while diffusion spreading acts as a dynamical coarse-graining for the different velocities of the fluid, which eventually leads to local thermalisation.  

\begin{figure}[t!]
\begin{center}
\includegraphics[width = 0.47\textwidth]{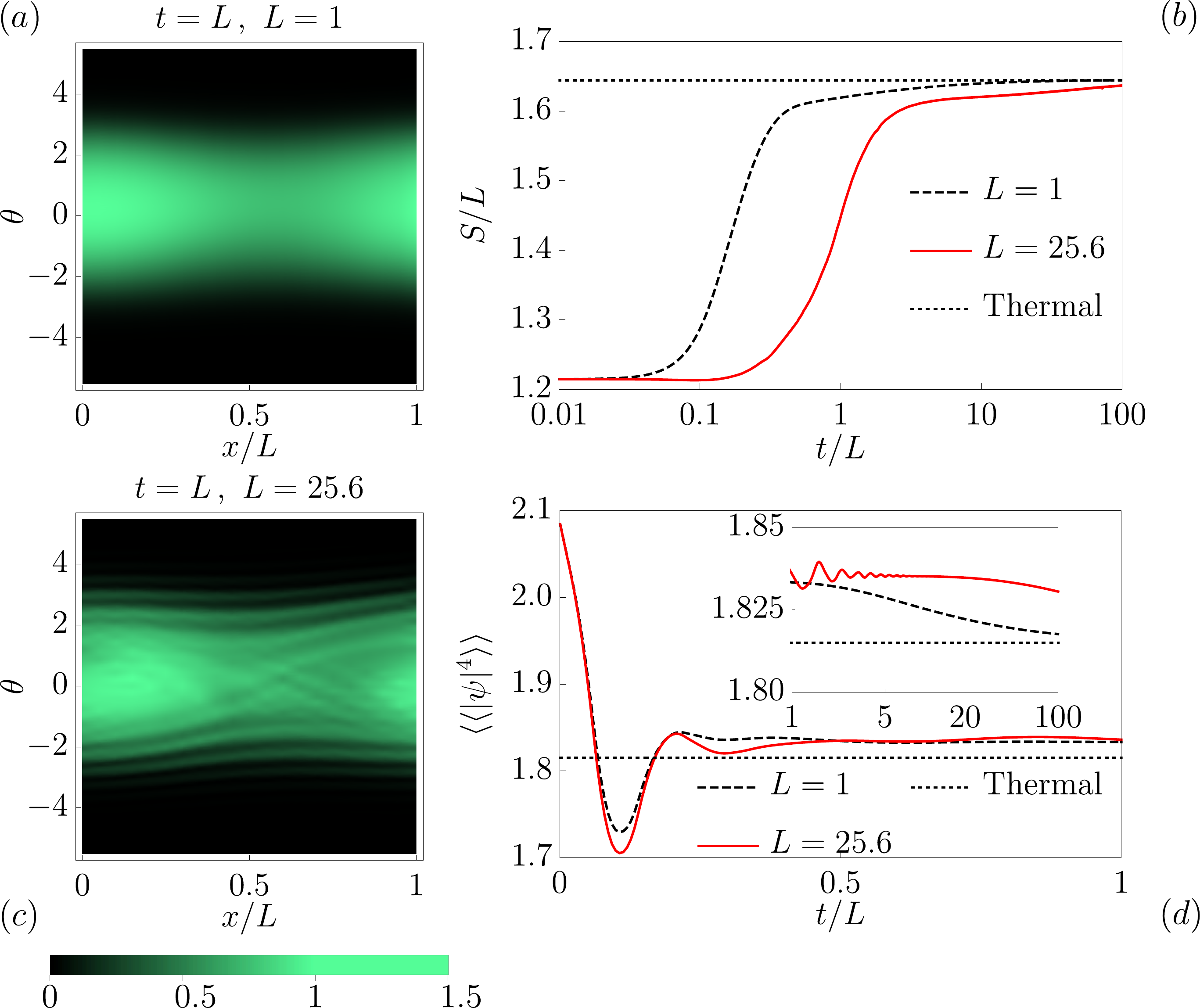}
\caption{ Left: Quasiparticle density $\rho_{\theta}(x,t)$ at $t=L$ for a Lieb-Liniger gas \eqref{eq:Htrap} inside the trap $V(x)= 1/2 (1- \cos 2 \pi x/L)$, in a circle with length $L=1$ (top) and $L=25.6$ (down) (see  text for other parameter's values). Right: Log-plot of the thermodynamic entropy density $\mathcal{S}(t)/L$ (top) converging to the thermal entropy for the given quench and of the integrated correlator $ \langle\langle |\psi^4 |  \rangle \rangle = \int_0^L dx \langle (\psi^\dagger \psi)^2 \rangle/L$. The inset is a log-plot to show the approach to the thermal value at late times.   }
\label{fig1}
\end{center}
\end{figure}
\paragraph{\textbf{ Lieb-Liniger gas in an external potential.}}
Our model of interest is the Lieb-Liniger model for a gas of bosons interacting with contact repulsion. The Hamiltonian is given by 
\begin{equation}\label{eq:HLL}
\hat{H}_{\rm LL}=\int d x\, \frac{\hbar^2}{2 m} \partial_x\psi^\dagger(x)\partial_x\psi(x)+g  \ \psi^\dagger(x)\psi^\dagger(x)\psi(x)\psi(x),
\end{equation}
with corresponding bosonic operators $[\psi^\dagger(x) ,\psi(y)] = \delta(x-y)$.
In the following we set the mass of the bosons such that  ${\hbar^2}/{2 m}  = 1$ and $g = c$. 
The model is known to describe state-of-the-art cold atom experiments \cite{Bloch2005,RevModPhys.80.885,Kinoshita1125,PhysRevLett.95.190406,Kinoshita2006,PhysRevLett.105.230402,PhysRevLett.100.090402,PhysRevLett.122.090601,2005.13646}
and also to be integrable  \cite{PhysRev.130.1605}.
As a consequence, all physical initial states~\cite{Kormos2013,PhysRevLett.111.100401,DeNardis2014,Kormos2014,PhysRevA.91.051602,Palmai2018} eventually relax to a GGE density matrix $\hat\varrho \sim  e^{- \sum_j \hat{Q}_j \beta_j}$, where the set of Lagrange multipliers $\beta_j$ is fixed by the initial condition. 

%It is now well established that the time evolution relaxes to a Generalised Gibbs Ensemble (GGE) for all physical initial conditions~\cite{Kormos2013,PhysRevLett.111.100401,DeNardis2014,Kormos2014,PhysRevA.91.051602,Palmai2018}. Namely, at large times all local expectations can be computed using the GGE with density matrix 
% $\hat\varrho \sim  e^{- \sum_j \hat{Q}_j \beta_j}$, where the set of Lagrange multipliers $\beta_j$ is fixed by the initial condition. 
%Under the assumption of relaxation to local GGE-like stationary states, its large scale dynamics can be accurately described via the so-called Generalised Hydrodynamics (GHD), introduced in the past years \cite{PhysRevX.6.041065, Bertini16} and which was recently shown to correctly predict an experimental realisation \cite{PhysRevLett.122.090601}. Contrarily to standard hydrodynamics which deals with finite numbers of conserved densities, GHD describes the dynamics of the densities of the infinite number of local and quasi-local conserved quantities $\hat{Q}_j = \int dx \hat{q}_j(x)$ characterising integrable systems. 

Its large scale dynamics can be accurately described via the so-called Generalised Hydrodynamics (GHD), introduced in the past years \cite{PhysRevX.6.041065, Bertini16} and the resulting predictions have even received an experimental confirmation \cite{PhysRevLett.122.090601}. 
In contrast with standard hydrodynamics, GHD is built on the infinite number of local and quasi-local conserved quantities $\hat{Q}_j = \int dx \, \hat{q}_j(x)$ characterising integrable systems. 
Equivalently, each stationary state of the system can be described by a set of stable quasiparticles with density $\rho_\theta$, in one-to-one correspondence with
the expectation values of conserved quantities $\int d\theta \rho_\theta \, h_{j, \theta} =  \langle \hat{q}_j \rangle$, with $h_{j, \theta}$, {the single-particle eigenvalues of each quasiparticle with rapidity $\theta$.}

%one can formulate the equilibrium and the dynamics of the model in terms of a set of stable quasipartices. Each stationary state of the system is encoded by the quasiparticle density $\rho_\theta$, in one-to-one correspondence with
%the expectation values of conserved quantities $\int d\theta \rho_\theta \, h_{j, \theta} =  \langle \hat{q}_j \rangle$.  
%The single-particle eigenvalues $h_{j, \theta}$ are model-dependent functions and the rapidity $\theta$ labels the  quasiparticle inner state, including its velocity. 
Given the derivative of the scattering shift between two quasiparticles with different rapidities $T_{\theta,\alpha}$, any single-particle function $h_{j,\theta}$ is dressed via the linear transformaton $h_j^{\rm dr} = (\mathbf{1}- T n)^{-1} \cdot h$, with $n_{\theta} = 2 \pi \rho_{\theta}/(k'_{\theta})$ the fermionic filling function and $(\mathbf{1}- T n)_{\theta,\alpha} = \delta_{\theta,\alpha} - T_{\theta,\alpha} n_\alpha$ the dressing matrix (here and in the following we denote the derivative respect to rapidities as $\partial_\theta h_\theta = h'_\theta$). The effective quasiparticle momentum $k_\theta$ and energy $\varepsilon_\theta$ are obtained by integrating the dressed derivatives: $k' = (\mathbf{1} - T n)^{-1}\cdot (k^{\mbox{\tiny bare}})'$ (and similarly for the energy). For the LL model, $T_{\theta,\alpha} = 2c/((\theta - \alpha)^2  + c^2)$, and $(k^{\mbox{\tiny bare}}_\theta)' = 1$, $(\varepsilon^{\mbox{\tiny bare}}_\theta)' = 2\theta$.
Because of the interactions, the effective energy and momentum of quasiparticles depend non-trivially on the local thermodynamic state via the dressing of excitations. 

Without external forces, the large scale evolution of any local density of conserved quantities can be written in terms of the evolution of the local density of quasiparticles $\rho_{\theta}(x,t)$. This involves a convective (Euler scale) and a diffusive part \cite{PhysRevLett.121.160603,10.21468/SciPostPhys.6.4.049,PhysRevLett.119.195301,vir1,PhysRevB.98.220303} and it reads  
$
\partial_t \rho_\theta  = - \partial_x(v^{\rm eff}_\theta \rho_\theta) + \partial_x( \mathfrak{D}_{\theta}^\alpha \partial_x \rho_\alpha),
$
where the repeated rapidity index $\alpha$ is summed over (summation over repeated indices is assumed in what follows, and includes an integration over velocities). 
The effective velocity $v^{\rm eff}_\theta = \varepsilon_\theta' /  k_\theta'$ gives the convective motion of densities \cite{PhysRevX.10.011054,2004.07113} while the non-diagonal diffusion kernel, which vanishes in non-interacting systems, describes both a diffusive spreading of the quasiparticle motion and a re-distribution of quasiparticle velocities. Both quantities are non-linear functional of the densities $\rho_\theta(x,t)$, fixed by the equation of state and by the explicit form of the diffusion kernel.

%In order to take into account the effect of an external potential, we consider the evolution given by the Hamiltonian 
The effect of an external trap is described by the following Hamiltonian
{
\begin{equation}\label{eq:Htrap}
\hat{H}_{\rm trap} = \hat{H}_{\rm LL} + \int dx \, \hat{V}(x)
\end{equation}
where the potential $\hat{V}(x) = V(x) \hat{q}_0(x)$ and the particle density $\hat{q}_0(x) = \psi^\dagger(x) \psi(x)$.}
The contribution of the trapping potential to the dynamics of the quasiparticle densities can be treated perturbatively in the degree of smoothness of the external potential $V(x)$. It was first shown in~\cite{doyon2017note} that
%, for this case of an external density field, 
at the Euler scale this amounts to adding the simple force term $(\partial_x V(x)) \partial_\theta \rho_\theta$. We here claim that the addition of this single term also gives the correct hydrodynamic equation up to diffusive scales:
\begin{equation}\label{eq:ghd_force}
\partial_t \rho_\theta  = - \partial_x(v^{\rm eff}_\theta \rho_\theta) + \partial_x( \mathfrak{D}_{\theta}^\alpha \partial_x \rho_\alpha) + (\partial_x V(x)) \partial_\theta \rho_\theta .
\end{equation}
{That is, this equation holds up to order $O(L^{-3})$ corrections, with $L$ the minimal length of spatial variations of the system and of the external potential $V(x)$.
From a scaling analysis, at this diffusive order possible corrections to Eq.~\eqref{eq:ghd_force} could come from three different sources: i) second derivative in the potential $\partial_x^2 V(x)$; ii) second power of its first derivative $(\partial_xV(x))^2$; and iii) mixed derivatives, $\partial_x V(x)\partial_x \rho_\alpha$. We now argue that none of these terms contributes.} 
First, we recall that the definitions of densities of charges $\hat{q}_j(x)$ are ambiguous by the addition of total derivative terms, and as shown in \cite{DeNardis2019}, we can make use of this freedom to enforce PT (parity and time) symmetry. {With PT symmetry, we can argue that terms proportional to $\partial_x^2 V$ are excluded in \eqref{eq:ghd_force}. Specifically, consider the Heisenberg time-evolution equation for any conserved density $\hat q_j(0)$, say at the origin. Taylor expanding the potential, $V(x) = V(0) + x V'(0) + O(x^2)$, we argue that the $O(x^2)$ does not contribute to Eq.~\eqref{eq:ghd_force}, as $\int\, dx \,x^2  [\hat{q}_0(x),\hat{q}_j(0)]$ has a vanishing expectation on any stationary state by PT symmetry.}  Second, no mixed derivatives appear as the charge density $\hat{q}_0(x)$ is \textit{ultra-local}. That is, in terms of the field $\psi(x)$ it does not contain any derivative, and as a consequence the number of particles does not generate any flow in space $[\int dx \hat{q}_0(x) ,q_i]=0$. This implies that diffusion is not modified by the presence of the external potential.
%This implies that the flow in the rapidity space $\partial_\theta \rho_\theta$ induced by the external potential has no diffusive corrections in real space $x$. 
{Third, there may be additional Fermi-golden-rule terms \cite{Mallayya2019,PhysRevB.101.180302,Durnin2020} leading to additional thermalisation effects. These arise from a second-order perturbation theory analysis with integrability-breaking perturbations. In our case, again concentrating on the origin, the leading integrability-breaking perturbation is  $ V'(0)B$ where $B=\int dx\,x\hat q_0(x)$. Fermi golden rule gives terms proportional to $(V'(0))^2$, of the time-integrated form $\int_{-t}^t ds \langle [B(s),\cdots]\rangle$, evaluated in a stationary, homogeneous state. As, in the present case, $B$ is the Galilean boost, $d B(s)/ds = P$ is the total momentum, thus, up to terms which vanish at large $t$, this is $\int_{-t}^t ds \,s\langle [P,\cdots]\rangle=0$ by homogeneity. Any integrable model with these properties is expected to obey \eqref{eq:ghd_force},  which is the focus of this letter. The above is not a full derivation, but we will address this, along with the general case, in an upcoming publication.}
%{Moreover, corrections due to dynamical violation of the local stationary state can lead at leading order to a Fermi-golden-rule term, whose first non-trivial order in derivatives is $(\partial_x V)^2$. This term can be shown to vanish for the case of external density field (see SM) due to the property that momentum $\hat{P} = -i/2 \int dx [\psi^\dagger(x) \partial_x \psi(x) - \partial_x \psi^\dagger (x) \psi(x)]$ is the current associated to the density $\hat{q}_0(x)$ and it is the only current operator conserved by \eqref{eq:HLL}.} Any integrable model with such property and in the presence of a density field is expected to be reproduced by \eqref{eq:ghd_force} but without momentum conservation and with different types of external potentials, the correct hydrodynamic equation is expected to display extra terms of order $O(L^{-2})$.

\paragraph{\textbf{Thermalisation via diffusion.}}
Canonical thermalisation to a Gibbs ensemble (GE) can only be reached at late times by breaking the conservation of all $ \hat{Q}_j$ except for the total energy $\hat{Q}_1 + \int dx \hat{V}(x)$ and the total particle number $\hat{Q}_0 = \int dx \psi^\dagger \psi$, and by ensuring that the dynamics is irreversible. 

{It can be shown that Gibbs ensembles are stationary solutions of the purely convective flow -- the first and third term on the right-hand side in \eqref{eq:ghd_force}~\cite{Note1, doyon2017note}. However, such an evolution is completely reversible \cite{Bulchandani2017,CauxCradle} and does not generate entropy. In practice, the root density $\rho_\theta$ is roughened by the convective evolution and cannot relax. Relaxation may happen in a weak sense as local observables may become stationary~\cite{Bulchandani2017,CauxCradle}~(see also Fig.~\ref{fig1}d). The determination of such a steady state is a non-trivial task. Physically, however, diffusion occurs, which modifies this picture. }
Diffusion -- the second term in \eqref{eq:ghd_force} -- increases entropy while preserving all conserved quantities. 
We will now show that the combined effect of diffusion together with the inhomogeneous potential $V(x)$ induces thermalisation.
%We will show that, in the presence of generic inhomogeneous potentials $V(x)$, the combination of Euler-scale breaking of conserved quantities and diffusive-scale entropy increase leads to canonical thermalisation.

% its only stationary solutions are those  now that we have a hydrodynamic equation \eqref{eq:ghd_force} valid  at the second order in the spatial derivative, we shall indeed show that, in the presence of generic inhomogenous potentials $V(x)$, it preserves total energy and particle number while breaking all other conserved quantities related to integrabilty, leading therefore to thermalisation to a standard Gibbs state.

 In order to proceed we shall first write down the form of the diffusion matrix $\mathfrak{D}$. This is given by $\mathfrak{D} = R^{-1} \widetilde{\mathfrak{D}} R$, where 
$
\widetilde{\mathfrak{D}}_{\theta,\alpha} = \delta_{\theta,\alpha} \int d\gamma \left( \frac{k'_\gamma}{k'_\alpha} \right)^2 W_{\gamma, \alpha} - W_{\theta, \alpha}
$
 and where the matrix of dressings $R_{\theta,\alpha} = 2\pi (\delta_{\theta,\alpha}- n_\theta T_{\theta,\alpha})/k_\theta' $.  The function $W_{\theta,\alpha} = 1/2 \left(2 \pi T^{\rm dr}_{\theta,\alpha}/k'_\theta \right)^2\rho_\theta (1- n_\theta) |v^{\rm eff}_\theta - v^{\rm eff}_\alpha|$, with  the dressed scattering kernel $T^{\rm dr} = (\mathbf{1}- T n)^{-1}\cdot T$. 
 The kernel  $\widetilde{\mathfrak{D}}_{\theta,\alpha}$ describes a Markov process of random momentum exchanges via 2-body collisions. However the total quasiparticle momentum is conserved, since $\int d\theta \mathfrak{D}_{\theta}^\alpha  = 1^\theta \mathfrak{D}_{\theta}^\alpha =0$, as it follows from $1^\theta [R^{-1}]_{\theta,\alpha} = (k'_\alpha)^2/(2\pi)$. The simple fact that the unit vector $1_\theta$ is the \textit{only} zero eigenvector of the diffusion operator $\mathfrak{D}$ ensures that the dynamics induced by Eq.~\eqref{eq:ghd_force} relaxes to a thermal ensemble. Namely the large time limit of $\rho_{\theta}(x,t)$ in  \eqref{eq:ghd_force} is such that, for any $j$, $\lim_{t \to \infty} \int d\theta \rho_{\theta}(x,t) h_{j,\theta}=\text{Tr} [\hat{q}_j(x) \hat{\varrho}]$ with the termal density matrix $\hat{\varrho}\propto\exp\left(-\beta(\hat{H}_\text{trap}-\mu\hat{Q}_0)\right)$. The final temperature  $\beta$ and chemical potential $\mu$ are fixed by the initial total density and energy.
%\begin{equation}\label{eq:GE}
%\lim_{t \to \infty} \int d\theta \rho_{\theta}(x,t) h_{j,\theta}  = \frac{{\rm Tr}[ \hat{q}_j(x) \exp\left(- \beta (\hat{H}_{\rm trap} - \mu  \hat{Q}_0\right)]}{Z},
%\end{equation}
%for any $j$,
%with $\beta$ and $\mu$ fixed by the initial total density $N$ and energy  $E$ and $Z ={\rm Tr}[   \exp\left(- \beta (\hat{H}_{\rm trap} - \mu  \hat{Q}_0\right)] $.

To proof this fact, first one notices that the total density  $N = \int dx \int d\theta \rho_\theta(x,t)$ and  total energy $E =\int dx \int d\theta \rho_\theta(x,t) (\varepsilon^{\rm bare}_\theta + V(x))$  are preserved by evolution \eqref{eq:ghd_force}. The latter can be easily shown using that $v^{\rm eff}_\theta =( k'_\theta)^{-1} [(\mathbf{1}- T n)^{-1}( \varepsilon^{\rm bare})']_\theta$. Then one proceed to notice that the diffusive term in \eqref{eq:ghd_force} produces an increase of the total thermodynamic entropy of the state $S= \int dx \sum_j \beta_j(x) \langle \hat{q}_j(x) \rangle - f$, with $f$ being the total free energy. The entropy increase on a generic GGE state was found in \cite{DeNardis2019,1912.08496} and it can be written  as 
\begin{equation}\label{eq:entropyincrease}
	\frac{d S(t)}{d t} = \frac{1}{2} \int d x\, \sum_{j,k}  (\partial_x \beta_j(x)) h_{j}^\theta (\mathfrak{D} C)_{\theta,\alpha} h_{k}^\alpha (\partial_x \beta_k(x) )
\end{equation}
 with the matrix of susceptibilities $[R C R^{-1}]_{\theta,\alpha}= \delta_{\theta,\alpha} \rho_\theta (1- n_\theta)$. 
As the only zero eigenvalue of the $\mathfrak{D}$ operator is the unit vector,  the only states where entropy does not increase are the ones where all $\beta_{j>0}$ are constant in space, namely any GGE states with an arbitrary spatially modulated chemical potential 
$\sim \exp( - \sum_{j=1}^m \hat{Q}_j \beta_j - \int dx \tilde\mu(x) \hat{q}_0(x))$, with all $\beta_j$ constants in space. 
However, for generic $m > 1$ and generic $\tilde{\mu}(x)$, these states are not invariant under the convective part of \eqref{eq:ghd_force} and therefore not stationary, leaving the thermal state $\tilde \mu = \mu - V(x)$ and $\beta_1 = \beta$ as the only stationary solution~\footnote{see supplementary material for a more detailed proof}.

%As the entropy increase must vanish on thermal states, this family includes the GE in Eq.~\eqref{eq:GE}, which corresponds to $\tilde \mu = \mu - V(x)$ and $\beta_1 = \beta$.
%However, for generic $m > 1$ and generic $\tilde{\mu}(x)$, these states are not invariant under the convective part of \eqref{eq:ghd_force} and therefore not stationary, see SM for a more detailed proof. We conclude then that the only stationary solution $\rho_\theta(x)$ are GEs \eqref{eq:GE}, namely when $m=1$ and $\tilde{\mu}(x) = \beta_1( \mu  - V(x))$. 
 
\paragraph{\textbf{Numerical solutions of the hydrodynamic equation.}}
The equation \eqref{eq:ghd_force} is a typical convection-diffusion equation.  A common way to numerically solve these types of equation is the Crank-Nicolson method \cite{cebeci2002convective}, which is based on the average of the forward and backword Euler step and is exact up to the second order in the time step.
We apply this method to Eq.~ \eqref{eq:ghd_force} and we solve the mid-point problem iteratively. 
More specifically, let us define the right hand side of \eqref{eq:ghd_force} as $
  \partial_t \rho_\theta(x,t) = F_\theta[\rho_\alpha (x,t)] 
$.
Note that the function $F_\theta$ is a non-linear functional of $\rho_\alpha(x, t)$. Let us denote as $\rho_\theta^{(k)}(x,{t+\Delta t})$ the approximation of $\rho_\theta(x,{t+\Delta t})$ after $k$ iterations. For the first iteration ($k = 1$) we use the forward Euler step
$
 \rho_\theta^{(1)}({t + \Delta t})(x, \theta) = \rho_\theta(x,t) + \Delta t F[\rho_\alpha(x,\theta)]
$. 
Then we obtain $\rho_\theta^{(k+1)}(x,{t + \Delta t})$ from $\rho_\theta^{(k)}(x,{t + \Delta t})$ using
$\rho_\theta^{(k+1)} = \rho_\theta(x,t) + \Delta t F_\theta\left[\frac{\rho_\alpha(x,t) + \rho_\alpha^{(k)}(x,{t + \Delta t})}{2}\right]$.
%\begin{equation}
%\rho_\theta^{(k+1)} = \rho_\theta(x,t) + \Delta t F_\theta\left[\frac{\rho_\alpha(x,t) + %\rho_\alpha^{(k)}(x,{t + \Delta t})}{2}\right].
%\end{equation}
Convergence is reached when 
$
 || \rho_\theta^{(k+1)}- \rho_\theta^{(k)}|| \ll \delta,
$
with $\delta$ some accuracy threshold.   The algorithm is very stable and preserves total energy with high precision. 
The role of the diffusion term in \eqref{eq:ghd_force} is to smoothen the functions $\rho_\theta(x,t)$ both in space and, primarily, in $\theta$ space, see Fig. \ref{fig1}, a fact that was previously noticed in \cite{Panfil2019} (see also  \cite{CauxCradle,CaoCradle}).
%
%The role of the diffusion term in \eqref{eq:ghd_force} is to smoothen the functions $\rho_\theta(x,t)$ both in space and, primarily, in $\theta$ space, see Fig. \ref{fig1}, a fact that was previously noticed in \cite{Panfil2019}. The smoothing in $\theta$ space corresponds to the lost of fine structures that are inevitably produced by the convective part together with the external force, see also \cite{CauxCradle,CaoCradle}. Such fine structures eventually lead to numerical instabilities, mostly due to the discretisation of the derivatives $\partial_\theta$. 
%
Therefore, the diffusion contribution not only is a necessary physical term to describe the hydrodynamics at order $\partial_x^2$, but it constitutes the proper regulator for any solution for the evolution of $\rho_\theta(x,t)$, by smoothing out all microscopic configurations.  However, whenever the ratio between the convective terms and the diffusive one is large (equivalently to a large Reynolds number in classical hydrodynamics) the profile of the fluid in $(x,\theta)$ space becomes rapidly varying, and numerical instabilities, mostly due to the discretisation of the derivatives $\partial_\theta$, inevitably appear.

\paragraph{\textbf{Results in the Lieb-Liniger gas.}}
%We are now in position to  numerically solve Eq. \eqref{eq:ghd_force}. 
We first study the following non-equilibrium settings: we consider periodic boundary conditions in $x \in [0,L]$. We prepare the gas in the thermal state  with inverse temperature $\beta=c=0.3$, with $c$ the coupling strength in \eqref{eq:HLL}, and with chemical potential given by  $\mu - V_0(x)$, with $\mu=2.5$ and $V_0= 2 \sin( 2 \pi x/L)$ and at $t=0^+$ we switch on the potential 
$
V(x)= 1/2(1- \cos( 2 \pi x/L))
$. In the equation \eqref{eq:ghd_force} one is free to rescale $x \to x L $ and $t \to t L$, this way the convective and the force term remain unaltered, but the diffusive part is rescaled by a factor $1/L$. %Therefore increasing $L$ reduces the strength of diffusion. 
%We study therefore the evolution with $L=1$ and $L=25.6$.  
In Figure \ref{fig1} we show the increase of the total entropy after the quench in the cases $L=1$ and $L=25.6$. 
%A rapid linear growth is followed by a slow drift towards the thermal value, which can be made even more pronounced by increasing $L$. 
For the large value of $L$ indeed we observe a pre-thermal plateau for the evolution of local observables, see Fig. \ref{fig1}.  {As discussed above, such a plateau corresponds to the stationary state attained by the purely convective evolution.} The slow drift can be interpreted as the effect of the spatial imbalance of modes with small diffusion constant, which produces a diffusive small entropy increase in \eqref{eq:entropyincrease}, and which is even more suppressed by larger values of $L$. We indeed stress that, even if the unit vector $1_\theta$ is the only zero eigenvector of the diffusion matrix, there exist a continuum of modes, namely the ones associated to quasiparticles with large momentum, with small diffusion constants. Such pre-thermal plateau was indeed also observed in numerical simulation of a trapped hard-rods gas in \cite{CaoCradle} and it is here therefore theoretically explained.
%: by increasing the smoothness of the gas diffusion is suppressed and increasing larger times are needed to observe the drift towards thermalisation. 

\begin{figure}[t!]
\begin{center}
\includegraphics[width = 0.49\textwidth]{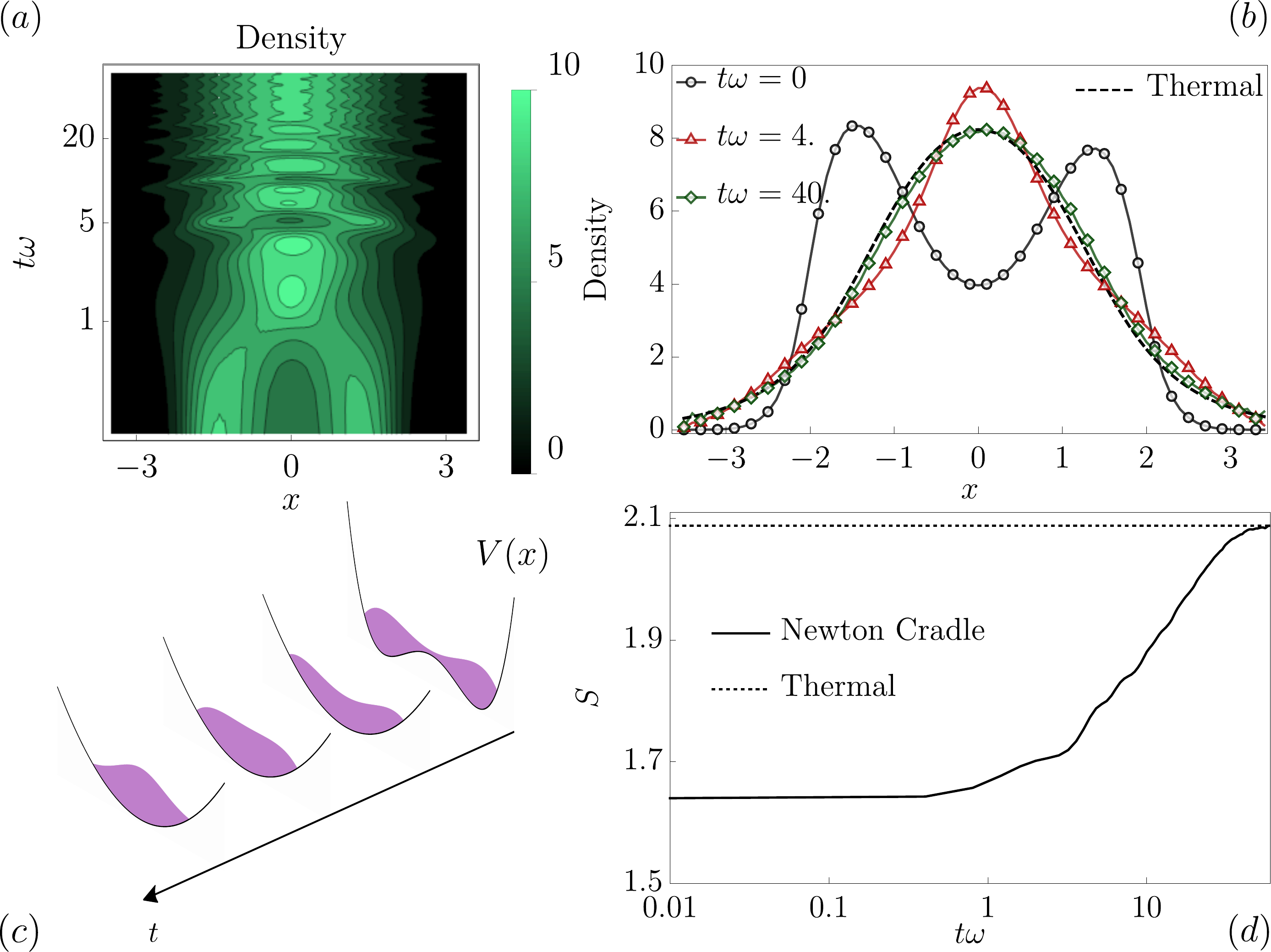}
\caption{ Study of the dynamics predicted by Eq. \eqref{eq:ghd_force} for a release of a cold atomic gas into a harmonic potential $V(x) = 2 x^2$ from a double well potential (see text for all parameters' values). (a) Plot of the density profile $\langle \hat{q}_0(x,t)\rangle$ as function of $t$ and $x$ and (b) at different times compared with its value in the the thermal Gibbs ensemble.  (d) Global thermodynamic entropy as function of time reaching the thermal value at late times. (c) Cartoon picture of the experimental setting.   }
\label{fig3}
\end{center}
\end{figure}

We then proceed to study a non-equilibrium setting which is more relevant for cold atoms experiments, see Fig. \ref{fig3}. We prepare a gas of bosons in a double-well potential as done in Ref. \cite{PhysRevLett.122.090601}. We consider an initial gas with $c=0.5$ at thermal equilibrium with temperature $T=10$ and $\mu=4$ inside the double-well trap $V_0(x) = x^4 - 2 x^2 + x^3/10$. We then release the gas in the harmonic trap $V(x) = \omega x^2/2$ with $\omega=4$. The initial density and temperature at the minima of the trap $V_0(x)$ is such that $c/n \sim O(10^{-2})$ and $  T/T_{\rm d}\sim O(10^{-1})$, with  quantum degeneracy temperature \cite{PhysRevLett.91.040403} $T_{\rm d} =  n^2$ (in units ${\hbar}^2/2m = 1$ and Boltzmann's constant $k_{\rm B} =1$), in order to be within typical experimental regimes. 
%By simulating the dynamics of the gas by \eqref{eq:ghd_force} we find that diffusion eventually suppresses the oscillations of the gas inside the harmonic trap and global entropy increases over around 40-50 oscillations inside the trap, to finally reach the value of its thermal equilibrium.  Interestingly the global entropy growth also display oscillations and its growth is not purely linear in time.  Such a slow thermalisation process explains why strict one-dimensional experimental settings have previously failed to observe thermalisation in purely one-dimensional settings, where the maximal experimental time included only 5-10 oscillations. 

By simulating the dynamics of the gas by \eqref{eq:ghd_force} we find that diffusion eventually suppresses the oscillations of the gas inside the harmonic trap and global entropy reaches thermal equilibrium over around 40-50 oscillations inside the trap.
%, to finally reach the value of its thermal equilibrium.  Interestingly the global entropy growth also display oscillations and its growth is not purely linear in time.  
Such a slow thermalisation process explains why strict one-dimensional experimental settings has previously failed to observe thermalisation, where the maximal experimental time included only 5-10 oscillations. 

\begin{figure}[t!]
\begin{center}
\includegraphics[width = 0.49\textwidth]{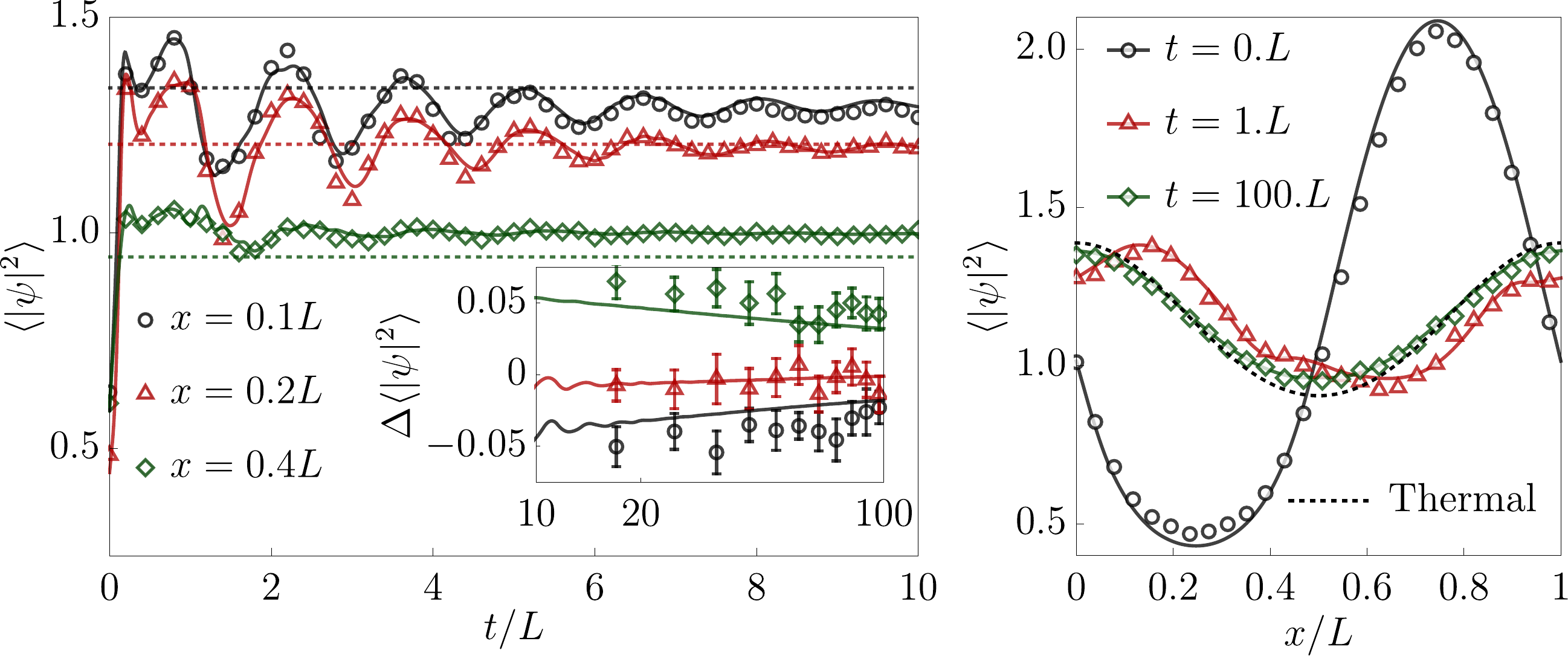}
\caption{ Comparison between the predictions of the hydrodynamic equation \eqref{eq:ghd_force} (continuous lines) and Monte-Carlo numerical simulation of the NLS model (dots with error bars) for the same quench as in Fig. \ref{fig1} with $L=25.6$. Left: evolution of the density profile $\langle |\psi|^2 \rangle  = \langle( \psi^\dagger  \psi) (x,t)\rangle $ at some given positions $x$ as function of time.  Inset shows the (diffusively) vanishing difference from the final thermal  expectation value as function of time $
\Delta\langle |\psi|^2 \rangle  \equiv \langle (\psi^\dagger  \psi) (x,t)\rangle  - \langle (\psi^\dagger  \psi) (x)\rangle _{\rm GE}$. Right: density profile for all $x \in [0,L]$ at different times. Microscopic simulations have been performed with the same method of Ref. \cite{bastianello2020generalised}}
\label{fig2}
\end{center}
\end{figure}

\paragraph{\textbf{Numerical simulations in the classical limit.}}
%In order to compare the predictions of the hydrodynamic equation \eqref{eq:ghd_force} with numerical simulation of the Lieb-Liniger gas we cannot rely on feasible methods.
%
Continuous quantum models are notoriously hard to  simulate with state--of--the--art tensor network techniques, despite recent progresses \cite{Verstraete2010,2006.01801}. We therefore benchmark the hydrodynamic equation \eqref{eq:ghd_force} with the classical limit of the Lieb-Liniger gas. Given the rescaling of the coupling $c = \hbar$, inverse temperature $\beta = \beta_{\rm NLS} \hbar$ {and the mass such that $\hbar^2/(2m) = 1$,}
in the limit $\hbar \to 0$  the quantum system is well described by the classical non-linear Schr\"odinger model (NLS)\cite{vecchio2020exact} (see also \cite{De_Luca_2016,10.21468/SciPostPhys.4.6.045,PhysRevB.101.245157}), where extensive numerical simulations can be carried \cite{doi:10.1080/09500340008232189,PhysRevLett.122.120401,PhysRevA.96.013623,PhysRevA.90.033611,PhysRevA.86.043626,PhysRevA.86.033626,PhysRevLett.122.120401}.
We numerically solve \eqref{eq:ghd_force} for the Lieb-Liniger gas and we extrapolated the behaviour at $\hbar =0$ to compare with numerical Monte-Carlo simulations of the NLS equation \cite{bastianello2020generalised}, see Fig. \ref{fig2}.
We consider the same quench of Fig. \ref{fig1} with $L=25.6$, in order to suppress extra effects to the dynamics due to corrections in \eqref{eq:ghd_force} with higher derivative terms.  Due to the large value of $L$ the thermalisation time-scales are large, see Fig. \ref{fig1}. Nevertheless we observe a small drift towards the thermal value for the profile of bosonic density, correctly reproduced by Eq. \eqref{eq:ghd_force}. 

%We numerically solve \eqref{eq:ghd_force} for the Lieb-Liniger gas with different values of $\hbar$  and we extrapolated the behaviour at $\hbar =0$ to compare with numerical Monte-Carlo simulations of the NLS equation \cite{bastianello2020generalised}, see Fig. \ref{fig2}. We numerically simulate the same quench of Fig. \ref{fig1} with $L=25.6$, in order to be deep inside the hydrodynamic regime and suppress extra effects to the dynamics due to corrections in \eqref{eq:ghd_force} with higher derivative terms.  Due to the large value of $L$ the thermalisation time-scales are large, see Fig. \ref{fig1}. Nevertheless we observe a small drift towards the thermal value for the profile of bosonic density,  correctly reproduced by Eq. \eqref{eq:ghd_force}. 
 
\paragraph{\textbf{Conclusion.}}
In this letter we have shown that the hydrodynamic equation \eqref{eq:ghd_force} including diffusion and external force term has Gibbs thermal states as the only stationary attractor states. This implies that interparticle diffusion is the leading mechanism for thermalisation in the presence of integrability breaking terms, and, as expected, no thermalisation is present in the non-interacting limit of hard-core bosons \cite{Mazza2014,PhysRevA.97.033609,PhysRevB.95.174303}, where no diffusion is present. This is in agreement with previous works on perturbed classical integrable models \cite{Lam2014,Goldfriend2019}, where it was noticed that, while chaotic behaviour is achieved at short times in many-body systems close to integrability, thermalisation requires to diffusively spread inside the classical phase-space, and it manifests itself on much longer time scales, see also \cite{Mazets2011} for a previous study on the Bose gas. We stress that this scenario evades KAM theorem \cite{kam2001}, where instead only a finite number of degrees of freedom are considered.  We here moreover show that the presence of a continuous number of modes in integrable models is responsible for a slow diffusion spreading as there is an infinite number of modes with arbitrarily small diffusion constant. On the other hand, thermalisation of local observables always occurs, as they {only depends on} a finite number of relevant modes, which leads to slow thermalisation drifts with diffusive finite-time corrections.  Our work gives a comprehensive explanation for the apparent lack of thermalisation on finite time-scales observed both in experimental \cite{Kinoshita2006} and numerical \cite{CaoCradle} settings and provide a quantitative prediction for the time-scales necessary to thermalise. In the near future we plan to study the hydrodynamic equation \eqref{eq:ghd_force} in the presence of generic external potential and for other models than Lieb-Liniger.  Moreover, while our work focused on a purely one-dimensional gas, one can now consider additions to \eqref{eq:ghd_force}  to account for different experimental effects as such as losses \cite{2006.03583}, dephasing noise \cite{bastianello2020generalised}, and crossover to 3-dimensional geometries \cite{Moller2020,2005.13546}. Finally, it is an interesting question for the future to relate the growth of thermodynamic entropy of Eq.~\eqref{eq:entropyincrease} to the well-known growth of entanglement entropy after quantum quenches  \cite{alba2017entanglement,Bertini2018,PhysRevX.7.031016,PhysRevB.101.094304,10.21468/SciPostPhys.7.1.005,2007.01286}.

\emph{\textbf{Acknowledgments}.}
A.B. acknowledges support from the European Research Council (ERC) under ERC Advanced grant 743032 DYNAMINT. J.D.N acknowledges King's College London for hospitality, Wojciech De Roeck for early collaborations on this work and numerous enlightening discussions. J.D.N. is supported by Research foundation Flanders (FWO). B.D. acknowledges discussions with Joseph Durnin, and related collaborations with M. J. Bhaseen.

\bibliography{biblio} 

%
%\section{Linearised GHD around thermal}
%To obtain the relaxation to the thermal state we choose $n = n_{\rm th} + \delta n$. By neglecting the term with $ \mathfrak{D}\partial_x^2 n_{\rm th}$ we have, using indices $i= (\theta,x)$
%\begin{equation}
%\partial_t n_\theta(x,t) = O_{\theta}^{\theta'} n_{\theta'}(x,t)
%\end{equation}
%with the linear operator computed on the thermal state and given by 
%\begin{align}
%O= - v^{\rm eff}_\theta \partial_x - \frac{\delta v^{\rm eff}}{\delta n} (\partial_x n_{\rm th}) \\+ \widetilde{\mathfrak{D}}_{\theta,\theta'}\partial_x^2 +[R\partial_x(R^{-1}  \widetilde{\mathfrak{D}})]_{\theta,\theta'}\partial_x +  V' \partial_\theta
%\end{align}

\newpage

\onecolumngrid
%\appendix
\newpage 

\appendix
\setcounter{equation}{0}
\setcounter{figure}{0}
\renewcommand{\thetable}{S\arabic{table}}
\renewcommand{\theequation}{S\arabic{equation}}
\renewcommand{\thefigure}{S\arabic{figure}}

\begin{center}
{\Large Supplementary Material \\ 
\titleinfo
}\\

\end{center}

\section{The thermal state as the only stationary state in the presence of diffusion}

Here we report a precise proof of what we briefly discussed in the main text, namely that the diffusive dynamics in the presence of an external trap necessarily leads to thermalisation. 
Let us consider the diffusive hydrodynamic equation Eq again. \eqref{eq:ghd_force}. For the sake of clarity, it is useful to introduce a global lengthscale $L$ to tune the inhomogeneity. Hence, we assume the potential has the scaling form $V(x)=U(xL^{-1})$ and we use a rescaled spatial coordinate $y=x L^{-1}$. After this rescaling, the hydrodynamic equation becomes

\be\label{eq_sm_diff}
\partial_t \rho_\theta=-L^{-1}\partial_y(v^\text{eff}_\theta \rho_\theta)+L^{-2}\partial_y(\mathfrak{D}^\alpha_\theta\partial_y \rho_\alpha)+L^{-1}(\partial_y V(x))\partial_\theta \rho_\theta+\mathcal{O}(L^{-3})
\ee

With $\mathcal{O}(L^{-3})$ terms coming from corrections beyond diffusion. We are ultimately interested in the regime of large $L$ with a suitable rescaling of time. Up to a timescale $t\sim\mathcal{O}(L^0)$, the system does not show any dynamics $\partial_t\rho_\theta\simeq 0$, since GGEs are stationary states of the local integrable model. Timescales $t\sim\mathcal{O}(L^1)$ display Eulerian dynamics, while finally $t\sim \mathcal{O}(L^2)$ will be affected by diffusion. Now, we show that the only stationary state that can be attained in the time window $\mathcal{O}\lesssim L^2\ll t\ll\mathcal{O}( L^3)$ is a thermal state in the local density approximation: since thermal states are expected to be stationary for the \emph{exact} dynamics, we conclude that the state will remain thermal also for $t\gg L^3$ where corrections beyond diffusion (not present in Eq. \eqref{eq_sm_diff}) cannot be neglected any longer.
As we already discussed in the main text, diffusion generally implies a growth in the Yang-Yang entropy: in the rescaled coordinates we have
\be
\frac{d  S}{d t}=L^{-2} \int d y\, \partial_y w_\theta (\mathfrak{D}C)_{\theta,\alpha} \partial_y w_\alpha
\ee
where we introduced the short-hand notation $w_\theta=\sum \beta_j(y) h_{j,\theta}$. The entropy grows on a timescale $t\sim \mathcal{O}(L^2)$: we assume a stationary state is reached, therefore the state must be such that $d  S/d t=0$. Since the only eigenvector of the diffusion matrix with zero eigenvalue is the constant function (in the rapidity space), we must have $\partial_\theta\partial_y w_\theta=0$, which leads to the general form
\be\label{eq_w}
w_\theta(x)=-\mu(y)+\nu(\theta)
\ee
with $\mu(y)$ an inhomogeneous local chemical potential (constant in the rapidity space) and $\nu(\theta)$ a function of only the rapidities, but constant in the coordinate space.
We now require that such an inhomogeneous GGE is also stationary for Eq. \eqref{eq_sm_diff}, i.e. we impose $\partial_t\rho_\theta=0$. 
Since the diffusion part in \eqref{eq_sm_diff} vanishes for the states of the form \eqref{eq_w}, we can focus on the stationarity with respect to the Eulerian dynamics. As explained in the main text (see also~\cite{doyon2017note}), this is enough to impose a thermal state.
In order to proceed, we first change the variable in the GHD equation from the root density to the generalised effective temperatures $w_\theta$.
To this end, we use the standard TBA results \cite{takahashi2005thermodynamics} connecting $w$ with the filling function
\be
\log(n_\theta^{-1}-1)=w_\theta-T_{\theta,\alpha} \log(1-n_\alpha)
\ee
Using this relation in Eq. \eqref{eq_sm_diff} and imposing the stationarity with  respect to the Eulerian part, standard manipulations leads to the simple equation

\be
v_\theta^{\text{eff}}(\partial_y w)^\text{dr}-(\partial_y U) (\partial_\theta w_\theta)^\text{dr}=0\,.
\ee
Now we use Eq. \eqref{eq_w} and notice that $(\partial_y w)^\text{dr}=-(\partial_y \mu(y))^\text{dr}=-\partial_y\mu(y)( k'_\theta)^\text{dr}$. In the last passage, we used that $\partial_y \mu(y)$ is constant in the rapidity space and therefore it can be moved out of the dressing. Using this last result, we reach the equality $(\partial_\theta \varepsilon_\theta^{\rm bare})(\partial_y \mu)=(\partial_y U) (\partial_\theta w_\theta)^\text{dr}$. Finally, using again that constants in the rapidity space commute with the dressing operation and inverting the dressing, we ultimately reach
the identity $\partial_\theta\varepsilon^\text{bare}\partial_y \mu=-\partial_y U \partial_\theta w_\theta$, which can be rewritten as
\be
\frac{\partial_\theta w_\theta}{\partial_\theta\varepsilon^\text{bare}}=-\frac{\partial_y \mu}{\partial_y U }\, .
\ee
The l.h.s. is a function only of the rapidities, while the r.h.s. is a function only of the spatial coordinate. The only possibility for the above identity to hold is that both ratios are constants. In particular $\partial_\theta w_\theta/\partial_\theta\varepsilon^\text{bare}=-\partial_y \mu/\partial_y U =\beta$: the constant $\beta$ can be immediately identified with the inverse temperature. Indeed, we can now write
\be
w_\theta=\beta(\varepsilon_\theta^\text{bare}+ U(y))+\text{const.}\, .
\ee
and finally conclude that the stationary state is thermal (within the local density approximation) with the undetermined constant playing the role of a global chemical potential.

 \end{document}